\documentclass{PoS}

\usepackage{epsf}
\usepackage{latexsym,amssymb,euscript}
\usepackage{amsmath}

\title{On a development of the phenomenological renormalization group}

\ShortTitle{On a development of the phenomenological renormalization group}

\author{\speaker{O.~Borisenko}\\
        Institute for Theoretical Physics, Kiev, Ukraine\\
        E-mail: \email{oleg@bitp.kiev.ua}}

\author{V.~Chelnokov\\
        Institute for Theoretical Physics, Kiev, Ukrainen\\
        E-mail: \email{chelnokov@bitp.kiev.ua}}

\author{V.~Kushnir\\
        Institute for Theoretical Physics, Kiev, Ukrainen\\
        E-mail: \email{vkushnir@bitp.kiev.ua}}

\abstract{ We propose a modification of the Nightingale renormalization group 
for lattice spin and gauge models by combining it with the cluster decimation 
approximation. Essential ingredients of our approach 
are: 1) exact calculation of the partition and correlation function on 
a finite lattice strip; 2) preservation of the mass gap or the second moment
correlation length, computed in the infinite strip length limit, on
each decimation step. The method is applied for studying general two and three dimensional 
Z(N) models. A perfect agreement with exact results (whenever available) is found. 
An extension of the method to models with a continuous symmetry is briefly discussed.}

\FullConference{31st International Symposium on Lattice Field Theory LATTICE 2013\\
                 July 29 - August 3, 2013\\
                 Mainz, Germany}

\begin{document}

\section{Introduction}

The Migdal-Kadanoff (MK) renormalization group (RG) \cite{migdal,kadanoff} 
had a great impact on the development of the real-space RG methods. Its 
conceptual and technical simplicity have probably been the main reason 
of the numerous applications of the MK RG ranging from two-dimensional ($2d$) 
classical spin models to the confinement problem in $4d$ gauge theories. In some 
special cases like $1d$ and hierarchical lattices the MK recursion relations (RR)
are precise and lead to an exact solutions. In other important cases the MK RG 
fails to correctly predict the phase diagram of the model, or gives the string tension 
which is too large \cite{ito}. Nevertheless, for a number of physically interesting systems 
the MK RG does provide qualitatively correct results which however are not very precise 
on the quantitative level. This fact has impelled people to look for a modification
of the original MK transformations which would result in a better quantitative predictions. 
Indeed, numerous such modifications have been developed in the past.

In the context relevant for this paper we would like to mention an approach to the RG 
transformations based on the cluster decimation approximation (CDA) \cite{CDA}.
The CDA utilizes the same restructuring of the lattice as in the conventional MK RG. 
The main idea is to preserve the free energy after each decimation step. 
This can be done approximately by computing exact free energies on small clusters with 
periodic boundary conditions (BC), e.g. $2\times 2$ and $1\times 1$. 
The equality of these free energies is used to establish RR for the effective coupling 
constant on the $1\times 1$ lattice. Then, this exact RR is used to approximate 
the {\it effective} coupling constant on new $L/2\times L/2$ lattice. On this way one gets 
a considerable improvement of the results over the standard MK RG. 

A different class of the RG is represented by a phenomenological renormalization group 
(PH RG) proposed by Nightingale \cite{nightingale} (for many applications of the PH RG, see 
\cite{nightingale1,yurishchev}). Originally, this approach utilizes 
the correlation lengths of the partially finite systems. Their scaling relations are then
interpreted as the RG equations. When expressed {\it via} eigenvalues of the transfer matrix 
on lattices with different sizes one obtains equations whose solutions approximate the critical 
temperature and various critical indices.

In the present contribution we develop an approach based on a combination of the PH RG with the CDA. 
If $m_r(L)$ is a mass gap of the finite system corresponding to a correlation function in representation 
$r$ and $b$ is a rescaling factor then in the vicinity of a critical point one has a scaling relation 
of the simple form $b m_r(L)\approx m_r(L/b)$. Basically, our approach can be described as follows.
\begin{itemize} 
 \item  The relations $b m_r(L)\approx m_r(L/b)$ are treated as a set of equations for 
new iterated couplings for the whole range of the bare coupling constants; 
\item In the spirit of the CDA we consider as a cluster the lattice strip $M^{(d-1)}\times L$, $L\to\infty$, 
where $d$ is dimension of the system. 
\end{itemize}
Alternatively, one can employ the second moment correlation length $\xi_2$ to obtain the phenomenological 
RG equations. This choice provides a good basis for Monte Carlo simulations on the square clusters. 
In the following we use both the mass gap and the second moment correlation length. 

This paper is organized as follows. In Section~2 we outline a general scheme 
of RG transformations on the example of the general $Z(N)$ spin models. 
Here, we also briefly discuss our approach to computation of partition and correlation functions 
on the finite lattice strips. In Section~3 we present some results of our study for two-and three-dimensional 
$Z(N)$ models. In Section~4 we summarise our results.

\section{Construction of RG transformations}

We work on a $d$-dimensional hypercubic lattice $\Lambda_0\in Z^d$
with lattice spacing $a=1$ and a linear extension $L$. $x=(x_1,x_2,...,x_d)$,
$x_i\in [0,L-1]$ denote the sites of the lattice, $l=(x,n)$ denote links 
and $e_n, n=(1,...,d)$ is a unit vector in $n$-th direction. We impose periodic BC. 
$\Lambda_k\in Z^d$ will denote lattice obtained after $k$ decimation steps with the 
extension $L_k=L/2^k$. Here we treat models only with interactions between nearest 
neighbours. The partition function (PF) of the general $Z(N)$ model can be written down as
\begin{equation}
Z(\Lambda_0, \{ t_k \} ) \  = \ \prod_{x\in \Lambda_0} 
\left ( \frac{1}{N} \sum_{s(x)=0}^{N-1} \right ) \ 
\prod_{l\in \Lambda_0} \ Q\left[ \{ t_k \}; s(x)-s(x+e_n)\right ] \ .
\label{PFZNdef}
\end{equation}
The most general $Z(N)$-symmetric Boltzmann weight is defined as 
\begin{equation}
Q\left[ \{ t_k \}; s \right ] \ = \ \sum_{k=0}^{N-1} \ t_k \ 
\exp\left[ \frac{2\pi i}{N} \ k \ s \right ] \ .
\label{Bweight1}
\end{equation}
The set of coupling constants $\{ t_k \}$ can be choosen to satisfy 
\begin{equation}
t_0=1 \ , \ 0 \leq t_k \leq 1 \ , \ t_k=t_{-k}=t_{k+N} \ .
\end{equation}
The two-point correlation function in the representation $r=1,\cdots ,N-1$ reads  
\begin{equation}
 \Gamma_r (\Lambda_0,\{ t_k \} ;R) \ = \ \langle \ 
\exp\left[ \frac{2\pi i}{N} \ r \ (s(x)-s(x+R)) \right ] \ \rangle \ .
\label{Gamma0}
\end{equation}
If all couplings are set to be equal, i.e. $t_1=t_2=\cdots = t_{N-1}$, one gets the standard Potts model. 
The coupling constants of the vector Potts model are given by    
\begin{equation}
t_k \ = \ C_k/C_0 \ , \ \ 
C_k \ = \ \sum_{m=0}^{N-1} \ \exp\left[ \beta\cos\frac{2\pi}{N} \ m + \frac{2\pi i}{N} \ m \ k \right ] 
\  = \  \sum_{s=-\infty}^{\infty} \ I_{Ns+k}(\beta)   \ .
\label{t_k0}
\end{equation}
Here, $I_k(x)$ is the modified Bessel function. 

Our goal is to present both partition and correlation functions on a decimated lattice 
in the form 
\begin{eqnarray}
Z(\Lambda_0, \{ t_k \} ) \  = \ A(\{ t_k \}) \ Z(\Lambda_1, \{ t_k^{(1)} \} )   \ ,  
\nonumber   \\
\Gamma_r (\Lambda_0,\{ t_k \} ;R) \ = \ G_r(\{ t_k \},R) \ \Gamma_r (\Lambda_1,\{ t_k^{(1)} \} ;R/2) \ 
\label{Z1G1}
\end{eqnarray}
with unchanged Boltzmann weight and a new set of couplings $\{ t_k^{(1)} \}$. Actually, any real-space 
RG amounts to a prescription of how to (approximately) compute, e.g. constant $A(\{ t_k \})$ and 
new couplings $\{ t_k^{(1)} \}$. We propose here to use the CDA of a certain type. We explain the main idea 
on the example of the two-dimensional lattice taking the rescaling factor $b=2$. Extension to higher 
dimensions and to other values of $b$ is straightforward. 

Let $\Lambda=(M\times L)$ be a strip of a $2d$ lattice with a width $M\ll L$ fixed and periodic 
BC in both directions. Define the free energy and the correlation function in the thermodynamic 
$L\to\infty$ limit as 
\begin{eqnarray}
F(M, \{ t_k \} ) \  &=& \ \log\lambda_0 (M,\{ t_k \} ) \ = \ 
\lim_{L\to\infty} \ \frac{1}{L} \ \log Z(M\times L, \{ t_k \} )   \ ,
\label{FM} \\
\Gamma_r (M,\{ t_k \} ;R) \ &=& \ \lim_{L\to\infty} \ \Gamma_r (M\times L,\{ t_k \} ;R) \ . 
\label{GM}
\end{eqnarray}
Suppose that $\Gamma_r (M,\{ t_k \} ;R)$ has the following very general form
\begin{equation}
\Gamma_r (M,\{ t_k \} ;R) \ = \ D_r(M,\{ t_k \},R) \ \left [ B_r(M,\{ t_k \}) \right ]^R \ ,
\label{formGM}
\end{equation}
where the function $B_r(M,\{ t_k \})$ describing the exponential decay is parametrized as 
\begin{equation}
B_r(M,\{ t_k \}) \ = \ \frac{\lambda_r (M,\{ t_k \} )}{\lambda_0 (M,\{ t_k \} )}
\label{BMpar}
\end{equation}
and the function $D_r(M,\{ t_k \},R)$ has utmost a power-like decay. Functions 
$\lambda_r$ can be considered, e.g. as the eigenvalues of the corresponding 
transfer matrices. Our basic idea is to present the correlation function $\Gamma_r (M,\{ t_k \} ;R)$ 
via the correlation function $\Gamma_r (M/2,\{ t_k^{(1)} \} ;R/2)$, calculated on the strip of 
the width $M/2$ with new couplings $\{ t_k^{(1)} \}$, in the form  
\begin{equation}
\Gamma_r (M,\{ t_k \} ;R) \ = \ \frac{D_r(M,\{ t_k \},R)}{D_r(M/2,\{ t_k^{(1)} \},R/2)} \ 
\Gamma_r (M/2,\{ t_k^{(1)} \} ;R/2) \ .
\label{RGtr1}
\end{equation}
Comparing (\ref{formGM}) with the last equation one concludes that (\ref{RGtr1}) holds if 
\begin{equation}
B_r^2(M,\{ t_k \}) \ = \ B_r(M/2,\{ t_k^{(1)} \}) 
\label{RGeq1}
\end{equation}
for all $r=1,\cdots ,N-1$. This system of $N-1$ equations determines new couplings 
$t_k^{(1)}$ on the lattice strip $(M/2,L/2)$. We use these exact relations to approximate 
the partition and the correlation functions on $\Lambda_0$ as  
\begin{eqnarray}
Z(\Lambda_0, \{ t_k \} ) \  &=& \ 
\left [ \frac{\lambda_0 (M,\{ t_k \} )}{\lambda_0^{(1/2)} (M/2,\{ t_k^{(1)} \} )} \right ]^{L^2/M} 
\  Z(\Lambda_1, \{ t_k^{(1)} \} ) \ , 
\label{RGpf} \\
\Gamma_r (\Lambda_0,\{ t_k \} ;R) \ &=& \  \frac{D_r(M,\{ t_k \},R)}{D_r(M/2,\{ t_k^{(1)} \},R/2)} 
\  \Gamma_r (\Lambda_1,\{ t_k^{(1)} \} ;R/2) \ .
\label{RGGM}
\end{eqnarray}
Equations (\ref{FM})-(\ref{GM}) and (\ref{RGpf})-(\ref{RGGM}) define our RG transformations. 
Renormalized coupling constants are computed from the equation (\ref{RGeq1}) which is, of course, 
the equation of the PH RG. When formulated in such a manner the PH RG becomes much more flexible tool, 
for it allows to perform many iterations. This is essential for the system with many coupling constants, 
be it original bare coupling constants or the constants generated during iterations. 
The second important gain is that within this approach one can compute not only critical characteristics 
of the system but also many thermodynamical functions in the whole range of the coupling constants.  
Also, instead of the mass gap preservation one can take any appropriate quantity whose scaling is known. 
For example, one could work with the second moment correlation length which is more suitable for the numerical 
Monte-Carlo computations. 

Similarly to any RG of this type, the main technical difficulty arises from the desire to obtain 
as precise results as possible. It is well known that the precision of the results systematically improves 
when the lattice strip width $M$ becomes larger and larger \cite{nightingale}. However, computation of the mass gap 
on a large strip requires diagonalization of big transfer matrices and this appears to be an important obstacle 
in the applications of this method. We have designed an original approach to the diagonalization problem 
details of which will be presented in \cite{sdarg}. Here we briefly outline two main ingredients of our approach.

\begin{enumerate}

\item The partition and correlation functions are computed in a dual formulation. {\it E.g.}, for $2d$ models 
one finds 
\begin{equation}
Z(\Lambda_0, \{ t_k \} ) \  = \ \sum_{r_n=0}^{N-1} \ \prod_{x\in \Lambda_0} 
\left ( \frac{1}{N} \sum_{s(x)=0}^{N-1} \right ) \ 
\prod_{l\in \Lambda_0} \  t_{s(x)-s(x+e_n)+\eta_n}  \ ,
\label{PFZNdual}
\end{equation}
where $\eta_n=r_n, n=1,2$ on a set of dual links which forms a closed path due to periodic boundary conditions, 
and $\eta_n=0$, otherwise. The transfer matrix is constructed in the dual formulation. The largest eigenvalue 
of the transfer matrix with $r_1=r_2=0$ gives the dominant contribution to the partition function while 
the largest eigenvalue of the transfer matrix with $r_1=r\ne 0,r_2=0$ gives the dominant contribution to the correlation 
function in the representation $r$ in the first direction. 

\item The transfer matrix itself is constructed not on the spin configurations but on coefficients describing 
evolution of all independent couplings. Such couplings are extracted after first direct summation over 
all spins from a given column in a strip. This approach automatically takes into account all lattice symmetries and 
significantly reduces the size of the matrix.

\end{enumerate} 

With this technics we have been able, for example, to compute the partition and the correlation functions for  
all standard Potts models up to $M=8$ in the presence of an external magnetic field, and for general 
$Z(N=4,5)$ models up to $M=8$ at zero magnetic field (to the best of our knoweledge, the partition function 
for all standard Potts models has been calculated exactly so far on the finite lattice strips 
for $M=2,3,4,5$, \cite{znstrip}).

\section{Application for $Z(N)$ models}  

Below we list some of the results obtained by applying above procedure to $2d$ $Z(N)$ spin models 
and to $3d$ $Z(N)$ lattice gauge theories. More applications together with technical details 
will be given in a forthcoming paper \cite{sdarg}.  

Our first example concerns the general $Z(4)$ spin model. The phase structure of this model is well known 
so that we can check both the validity and the precision of our approach. Fixed points are extracted from the 
solution of the recursion relations and the critical lines can be obtained after just several transformations since 
the RG iterations converge rather fast. Critical index $\nu$ is derived in a usual way by linearising the RG equations 
around fixed points. In this case we find 4 fixed points corresponding to: 1) the standard Potts model ($t_1=t_2$), 
2) the vector model ($t_2=t_1^2$), 3) and 4) models defined on lines $t_1=0$ and $t_2=1$. Our results are summarised 
in Table~\ref{tbl:crittnu4} (results for the line $t_2=1$ reproduce the ones for $t_1=0$). 

\begin{table}[ht]
\centering
\begin{tabular}{||c||c|c||c|c||c|c||}
\hline
$M_1\rightarrow M_2$ & $t_{s}$ & $\nu_{s}$ & $t_{v}$ & $\nu_{v}$ & $t_2 (t_1=0)$ & $\nu$ \\
\hline
  Exact & 1/3 & 2/3 & $1/(1 + \sqrt{2})$ & 1 & $1/(1 + \sqrt{2})$ & 1 \\
 $4\rightarrow2$ & 0.321807 & 0.730859 & 0.402491 & 0.934196 & 0.402491 & 0.934217 \\
 $6\rightarrow2$ & 0.326477 & 0.734682 & 0.406834 & 0.94718 & 0.406835 & 0.947154 \\ 
 $6\rightarrow4$ & 0.330436 & 0.741442 & 0.411239 & 0.970341 & 0.411239 & 0.970634 \\ 
 $8\rightarrow2$ & 0.328659 & 0.736383 & 0.408962 & 0.95484 & 0.408962 & 0.954708 \\
 $8\rightarrow4$ & 0.331406 & 0.743497 & 0.412259 & 0.97711 & 0.41226 & 0.977542 \\
 $8\rightarrow6$ & 0.332276 & 0.746328 & 0.41329 & 0.98673 & 0.413294 & 0.987379 \\
\hline
\end{tabular}
\caption{Critical coupling and critical index $\nu$ for $Z(4)$ model 
obtained from iterations $ M_1 \times L\rightarrow M_2\times L$. 
Here subscript ``s'' stands for the standard model and ``v'' - for the vector model.}
\label{tbl:crittnu4}
\end{table}
In our next example we use the second moment correlation length as a quantity which we preserve during 
RG steps. In this case we consider a square cluster of size $L\times L$ and the RG of the form 
$L\to L/2$. Table~\ref{rg_xi2} summarises results of our Monte-Carlo simulations 
for several Potts models. Subscript $e$ refers to exact values.  

\begin{table}[ht] 
\centering
\begin{tabular}{|l|c|c|c|c|c|}
\hline
N & L & $\beta_c$ & $\beta_{c(e)}$ & $\nu$ & $\nu_{(e)}$ \\
\hline
\hline
2 & 16 & 0.441905 & & 1.04733 & \\
 & 32 & 0.440965 & & 1.01295 &  \\
 & 64 & 0.440664 & 0.440687 & 0.998986 & 1.0\\
\hline
3 & 8 & 0.33703 & & 0.971028 & \\
 & 16 & 0.33531 & & 0.887692 & \\
 & 32 & 0.33505 & & 0.857852 & \\
 & 64 & 0.3350186 & 0.335018 & 0.849067 & 5/6\\
\hline
5 & 16 & 0.234663 & & - & - \\
 & 32 & 0.234726 & & - & - \\
 & 64 & 0.2348156 & 0.234872 & - & - \\
\hline
13 & 8 & 0.1165 & & - & -\\
 & 16 & 0.117 & & - & - \\
 & 32 & 0.1175 & 0.117482 & - & - \\
\hline
\end{tabular}
\caption{Critical coupling $\beta_c$ and critical exponent $\nu$ for $Z(N), N=2, 3, 5, 13$}
\label{rg_xi2}
\end{table}

We have also studied the general $3d$ $Z(N)$ gauge models for various $N$. In this case we also used 
duality transformations and performed RG iterations with the smallest cluster $2\times 2\times L$. 
In Table~\ref{rg_results} we present the estimates of the critical points
$\beta_{c}$, both in standard Potts gauge models and in vector gauge models 
and compare them with the results of Monte Carlo numerical simulations. 

\begin{table}[ht]
\begin{center}
\begin{tabular}{|c|c|c|c|c|c|}
\hline
    & \multicolumn{2}{|c|}{Potts model} & \multicolumn{3}{|c|}{Vector model}\\
\cline{2-6}
$N$ & $\beta_{\rm c}$ & $\beta_{\rm c}^{\rm MC}$ & $\beta_{\rm c}$ 
& $\beta_{\rm c}^{\rm MC}$ & $\nu$ \\ 
\hline
2  & 0.77706 & 0.761414(2) & 0.77706 & 0.761395(4) & 0.616656 \\
3  & 1.17186 & 1.084314(8) & 1.17186 & 1.0844(2)   & -        \\
4  & 1.34363 & 1.288239(5) & 1.55411 & 1.52276(4)  & 0.616657 \\
5  & 1.50331 & 1.438361(4) & 2.17896 & 2.17961(10) & 0.692226 \\
6  & 1.62881 & 1.557385(4) & 2.99296 & 3.00683(7)  & 0.699208 \\
8  & 1.81941 & 1.740360(6) & 5.09472 & 5.12829(13) & 0.699583 \\
\hline
\end{tabular}     
\caption{$3d$ $Z(N)$ standard Potts models: $\beta_{c}$ from the PH RG 
(column two) and from the Monte Carlo simulations of Ref.~\cite{bazavov} 
(column three). $3d$ $Z(N)$ vector models: $\beta_{c}$ from the PH RG (column~4) and from 
the Monte Carlo simulations of Ref.~\cite{zn3d} (column five); critical index $\nu$
from the PH RG.}
\end{center}
\label{rg_results}
\end{table}

\section{Summary and perspectives} 

In this paper we considered a simple modification of the PH RG by combining it with the CDA. 
Within these frameworks we have derived exact representations for the partition and correlation 
functions on the decimated lattice. Recursion relations are derived from the requirement of 
the preservation  of the mass gap (or the second moment correlation length) of the system 
for each representation of the correlation function on every RG iteration. 
We then presented results of application of this 
approach to various two- and three-dimensional $Z(N)$ models. One sees from the Tables given in 
Section~3 that RG equations obtained even on small lattice clusters give very reasonable 
approximations to exact or numerical values.   
Clearly, the above procedure can be easily extended to models with continuous symmetry
like the $XY$ model, the principal chiral model and the $O(N)$ sigma model as well as to gauge theories. 
The details of the corresponding constructions will be given elsewhere.

\end{document}